\documentclass[aps,prl,twocolumn,nofootinbib,a4paper,amsfonts,amssymb,amsmath,floatfix,superscriptaddress]{revtex4-2}

\usepackage{hyperref}
\usepackage{multirow}
\usepackage{mathrsfs,amsmath} 
\usepackage{verbatim}
\usepackage{amsmath}
\usepackage{amsfonts}
\usepackage{amssymb}
\usepackage{amsthm}
\usepackage{bm}
\usepackage{xparse}
\usepackage{graphicx}
\usepackage{xcolor}
\usepackage{textcase}
\usepackage{url}
\usepackage{lipsum}
\usepackage{appendix}
\usepackage{dsfont}
\usepackage{epstopdf}
\usepackage{footnote}
\usepackage[caption=false]{subfig}
\usepackage{float}
\usepackage{subfig}
\usepackage{soul}
\usepackage{xcolor}
\newcommand{\ket}[1]{| {#1} \rangle} 
\DeclareDocumentCommand{\Tr}{m m O{\big}}{{\rm Tr}_{\:\!{#1}}#3({#2}#3)}

\begin{document}
\title{ Quantum coherence in networks}
\author{Fatemeh Bibak}
\affiliation{University of Vienna, Faculty of Physics, Vienna Center for Quantum
Science and Technology, Boltzmanngasse 5, 1090 Vienna, Austria}
\affiliation{Institute for Quantum Optics and Quantum Information (IQOQI),
Austrian Academy of Sciences, Boltzmanngasse 3,
A-1090 Vienna, Austria
}
\author{Flavio Del Santo}
\affiliation{Group of Applied Physics, University of Geneva, 1211 Geneva 4, Switzerland}
\affiliation{Constructor University, Geneva, Switzerland }
\author{Borivoje Daki\'c}
\affiliation{University of Vienna, Faculty of Physics, Vienna Center for Quantum
Science and Technology, Boltzmanngasse 5, 1090 Vienna, Austria}
\affiliation{Institute for Quantum Optics and Quantum Information (IQOQI),
Austrian Academy of Sciences, Boltzmanngasse 3,
A-1090 Vienna, Austria
}

\date{\today}

\begin{abstract}
From a quantum information perspective, verifying quantum coherence in a quantum experiment typically requires adjusting measurement settings or changing inputs. A paradigmatic example is that of a double-slit experiment, where observing the interference pattern on the screen in a series of experimental settings where one, the other, and both slits are open unambiguously proves quantum coherence. Here we show that this is not necessary by verifying quantum coherence in a network scenario without the need for inputs. We show that there exist probability distributions for joint outcomes of three parties in a triangular network with independent sources that cannot be replicated using classical resources. Furthermore, we generalize our results to $n$-party networks and show that the discrepancy between correlations in classical and quantum networks increases with the number of parties. To this end, we derive nonlinear inequalities that are satisfied by classical correlations and find quantum states that violate them.
\end{abstract}

\maketitle

\section{Introduction}

One of the main features that distinguishes quantum physics from its classical counterpart is the superposition principle, i.e., the fact that linear combinations of quantum states form another valid state. Since the early days of quantum information science, the power of coherent quantum superposition has been exploited to achieve tasks fundamentally unattainable in classical scenarios, such as secure cryptography \cite{bennett2014quantum}, or algorithms showing an exponential advantage over their classical analogue  \cite{deutsch1992rapid, shor1994algorithms}. Moreover, it was more recently shown that quantum coherence of order \cite{feix2015quantum, jia2019causal} and of direction of communication \cite{guerin2016exponential,  chiribella2013quantum, chiribella2018indefinite} prompts an exponential reduction of communication complexity, leads to two-way \cite{del2018two} and multi-way signaling \cite{horvat2021interference} with a single quantum particle, enhances the information acquisition speed \cite{horvat2021quantum}, and allows to access information via quantum indistinguishability \cite{horvat2023accessing}, etc.

Detecting quantum coherence, however, requires indirect measurements that allow the observation of interference phenomena. 
Formally, this is due to the fact that, while classical probabilities are added up directly, quantum probabilities correspond to the modulus square of the added amplitudes. It ought to be stressed that to reveal quantum superposition of single quantum systems, for instance in a standard double-slit experiment (or, equivalently, in a Mach-Zehnder interferometer), one needs always to compare different and incompatible scenarios. Label with ``0'' or ``1'' the setup where one of the slits is  open or closed, respectively.  This leads to four possible configurations, ``00'', ``01'', ``10'', or ``11'',  where the first (second) bit indicates the setup of the first (second) slit. The interference term is then given by \cite{dakic2014density,sorkin1994}:
\begin{equation}\label{interf}
I:= \sum_{i,j=0}^1 (-1)^{i\oplus j} p_{ij},
\end{equation}
where $p_{ij}$ is the conditional probability to detect the particle at a particular location on the screen after the slits, given the configuration $ij$. For  classical particles it is straightforward to show that $I=0$. However, in quantum mechanics $I$ can be non-zero. This and other standard tests that demonstrate the advantage of quantum resources over their classical counterparts, such as in tests of quantum nonlocality or contextuality, require a change of measurement settings or inputs in modern game-theoretical language \cite{scarani2014bell}. This is also the case in testing coherence via tuning the interferometric phase (e.g. the phase-shift in Mach-Zehnder interferometer) \cite{del2020coherence, horvat2021interference}. 

In this letter we put forward a protocol that allows to witness quantum coherence without the need to change inputs. Motivated by the triangle network in the search of new forms of nonlocality without inputs, we notice that our setup is a single-particle modification of the triangle network which has  recently been widely studied in the context of genuine network nonlocality \cite{renou2019genuine, boreiri2023towards,tavakoli2022bell, sekatski2023partial, 
 boreiri2023noise}. There, in a similar fashion to what we present here, it is possible to detect Bell nonlocality without inputs at the cost of assuming independence of the sources that emit quantum particles. We show that in a three-party scenario (arranged on a triangle network) with independent sources the parties cannot return outputs which fulfill a certain  probability distribution if they only use classical resources. However, this can be achieved if  they exploit quantum coherence. 
We also derive a nonlinear inequality that separates correlations originating from classical sources from those using quantum coherence. Finally, we extend our findings to networks with an arbitrary number of parties. With extensive numerical analysis, we demonstrate that the discrepancy between classical and quantum correlations increases with the number of parties, reaching its maximum value in the asymptotic limit. Throughout the paper presentation, for the sake of simplicity, we shall assume that the particles in question are single photons. Nevertheless, an analogous calculation can be performed for fermions or distinguishable particles, with no appreciable difference. 

\section{Triangle network for quantum coherence}

We present a protocol that can distinguish between classical scenarios and those exhibiting quantum coherence, which does not necessitate a change of settings (or inputs) between different runs.
As illustrated in Fig. \ref{fig:threepartiesclassical}, the protocol features three parties, Alice (A), Bob (B) and Claire (C) each of whom is positioned on a different vertex of a triangle.  On every edge, i.e. in between each pair of parties, there is a source  --each labeled $S_i$, with $i=1,2,3$--  of physical systems or \textit{information carriers} (i.e. single-particle), which can be either classical or quantum in nature. Each source can send its corresponding particle only to the parties lying at the end of the respective edge (see Fig.~\ref{fig:threepartiesclassical}). The parties, possibly making use of the resources that they receive from the adjacent sources, perform local operations (e.g., measurements) and each return a binary outcome, $a$, $b$, and $c$, respectively. Note that A, B, and C do not change measurement setting druing the entire experiment.

The task is to return the probability distribution:
\begin{align}
\label{predicate}
p(a,b,c|a\oplus b \oplus c & = 0)\neq 0 \ \ \&  \nonumber \\
p(a,b,c|a\oplus b \oplus c & = 1)= 0 \ \ \ \forall \,a, b, c
\end{align}
constrained with the following assumptions:
\begin{itemize}
    \item[] (i) \textit{Single information carriers} -- each source produces one information carrier (i.e., a particle) in a single experimental run; 
    \item[] (ii) \textit{Independent sources} -- the probability distribution associated to the emission of different sources are uncorrelated.
\end{itemize}

Note that assumption (i) can be operationally verified by a simple inspection (e.g. by measuring the path of the particle) that only a single information carrier (particle) is present in the two output beams of each source. 

Furthermore, we will consider only postselected scenarios in which each party detects a single particle. This is to prevent two particles interacting at detection stations A, B or C.

\subsection{Classical network}

Let us now consider the most general classical network, i.e., when the information carriers can only be probabilistically directed from the source to one or the other measurement station (intuitively, this corresponds to the probabilistic distribution of trajectories of well-localized classical particles). Each independent source has a certain probability to send the particle either to the left or to the right: $S_1$ sends the particle to Alice with probability $\lambda_1$ or to Bob with probability $\lambda_2=1-\lambda_1$,  $S_2$  to Bob with probability $\nu_1$ or to Claire with probability $\nu_2=1-\nu_1$, and  $S_3$ to Claire with probability $\mu_1$ or to Alice with probability $\mu_2=1-\mu_1$ (see Fig. \ref{fig:threepartiesclassical}).
After post-selecting to a single particle per party, Alice, Bob, and Claire observe the following joint probability distribution:
\begin{figure}[h!]
    \centering
    \includegraphics[width=0.7\linewidth]{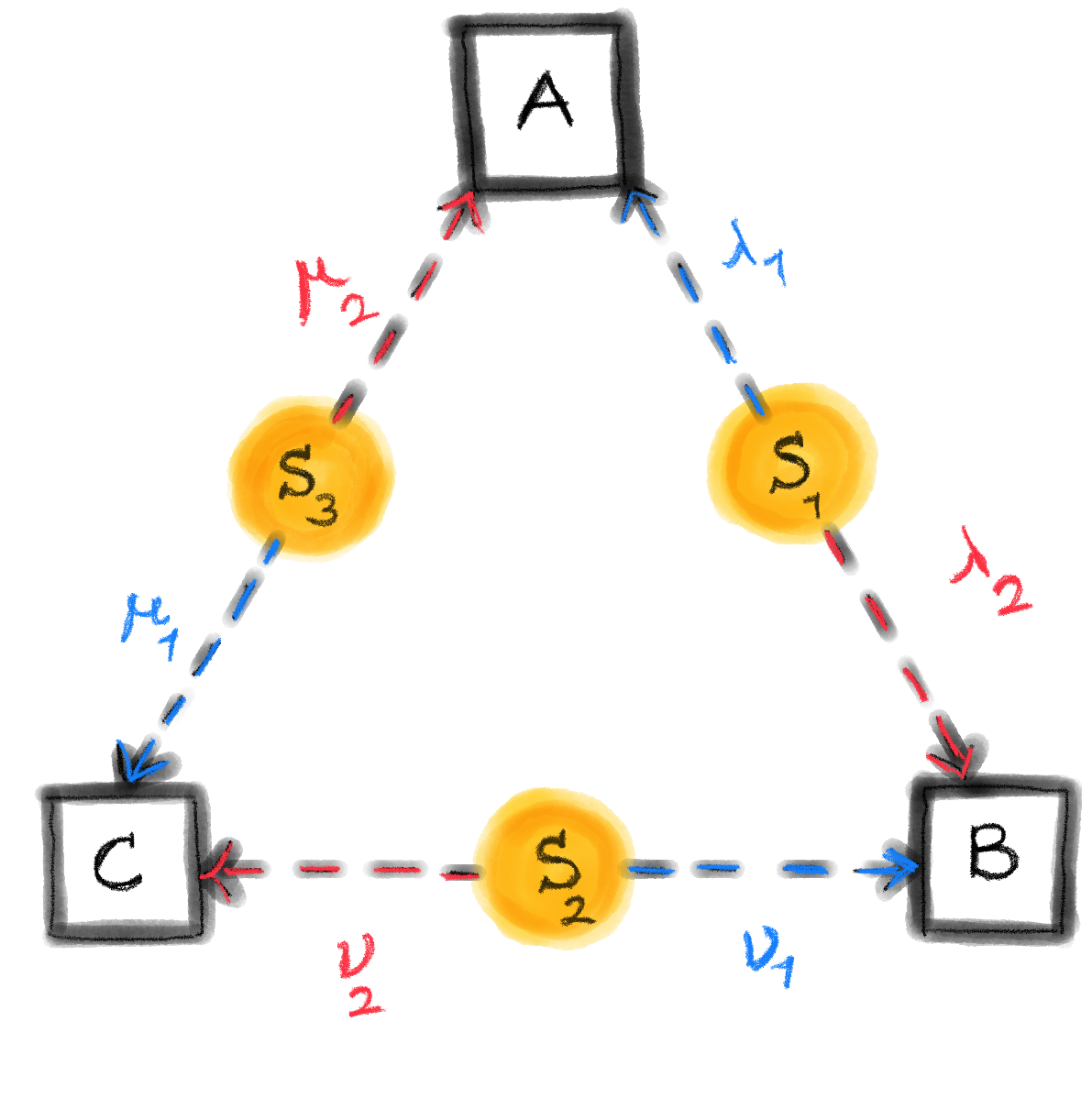}
    \caption{Three-party classical network. In the triangle scenario with classical sources, the individual particles are directed with classical probabilities to the measurement stations at the vertices of the triangle. For example, source $S_1$ sends the particle to Alice's lab with probability $\lambda_1$ and to Bob's lab with probability $\lambda_2$. Here we consider only postelected scenario, i.e., where only one particle ends up in each lab (see main text for details). In such case, there are only two possibilities, all sources send their particles clockwise or counterclockwise. }
    \label{fig:threepartiesclassical}
\end{figure}
\begin{align}
    p(a,b,c)=&(\frac{\lambda_1\mu_1\nu_1}{\pi_{111}}p(a|S_1)p(b|S_2)p(c|S_3)+\nonumber\\&\frac{\lambda_2\mu_2\nu_2}{\pi_{111}}p(a|S_3)p(b|S_1)p(c|S_2)),
    \label{eq:clprob3}
\end{align}
where $\pi_{111} = \lambda_1 \mu_1 \nu_1 + \lambda_2 \mu_2 \nu_2$ is the total probability for a single particle detection, and $p(a|S_1(S_3))$ is the probability that Alice obtains outcome $a$ given that she received the particle from $S_1$ ($S_3$). The same notation holds for Bob's and Claire's outcomes.
 This probability distribution can be understood as a classical (convex) mixture of the probabilities for the case where all sources emit particles either towards the "left" or the "right" from them. This postselection corresponds to the sending the three particles in a "clockwise" or "counterclockwise" direction within the network, respectively. 

The condition in Eq. \eqref{predicate}, stipulates that the parity of the three outcomes $a$, $b$, $c$ is never odd.
This and Eq.~\eqref{eq:clprob3} lead to the conjunction of the following set of conditions:
\begin{align}
\label{c11}\{p(a=1|S_1)\,\text{or} \,p(b=1|S_2)\,\text{or}\,p(c=1|S_3)=0 \}\nonumber \ &\&&\nonumber \\\{p(a=1|S_3)\,\text{or}\,p(b=1|S_1)\,\text{or}\,p(c=1|S_2)=0\};\\ \nonumber \\
\label{c22}\{p(a=0|S_1)\,\text{or}\,p(b=0|S_2)\,\text{or}\,p(c=1|S_3)=0\}\nonumber \ &\&&\nonumber \\\{p(a=0|S_3)\,\text{or}\,p(b=0|S_1)\,\text{or}\,p(c=1|S_2)=0\};\\ \nonumber \\
\label{c33}\{p(a=0|S_1)\,\text{or}\,p(b=1|S_2)\,\text{or}\,p(c=0|S_3)=0\} \nonumber \ &\&&\nonumber \\\{p(a=0|S_3)\,\text{or}\,p(b=1|S_1)\,\text{or}\,p(c=0|S_2)=0\};\\ \nonumber \\
\label{c44}\{p(a=1|S_1)\,\text{or}\,p(b=0|S_2)\,\text{or}\,p(c=0|S_3)=0\} \nonumber\ &\&&\nonumber \\\{p(a=1|S_3)\,\text{or}\,p(b=0|S_1)\,\text{or}\,p(c=0|S_2)=0\}.
\end{align}
It is straightforward to demonstrate that conjunction of all above conditions (\ref{c11}-\ref{c44}) results in a contradiction with the condition $p(a, b, c|a\oplus b \oplus c = 0)\neq0$ given in Eq.~\eqref{predicate}. To illustrate this, let us fix the following probabilities: \( p(a=1|S_1)=0 \) and \( p(b=1|S_1)=0 \). This way, condition \eqref{c11} is satisfied. Next, we fix \( p(b=0|S_2)=0 \) and \( p(a=0|S_3)=0 \) to meet condition \eqref{c22}. To satisfy condition \eqref{c33}, we must have \( p(c=0|S_3)=0 \). Finally, to meet condition \eqref{c44}, we have no other choice but \( p(c=0|S_2)=0 \). This results in a contradiction with Eq.\eqref{predicate}, as both \( p(c=0|S_2)=0 \) and \( p(c=0|S_3)=0 \), leading to \( p(1,1,0)=0 \) and \( p(0,0,0)=0 \). Using the same logic, one can disregard other combinations with zero probabilities that satisfy conditions (\ref{c11}-\ref{c44}).  

This proves that classical networks are incompatible with the condition~\eqref{predicate}.
\subsection{Quantum network}
\begin{figure}[h!]
    \centering
    \includegraphics[width=0.7\linewidth]{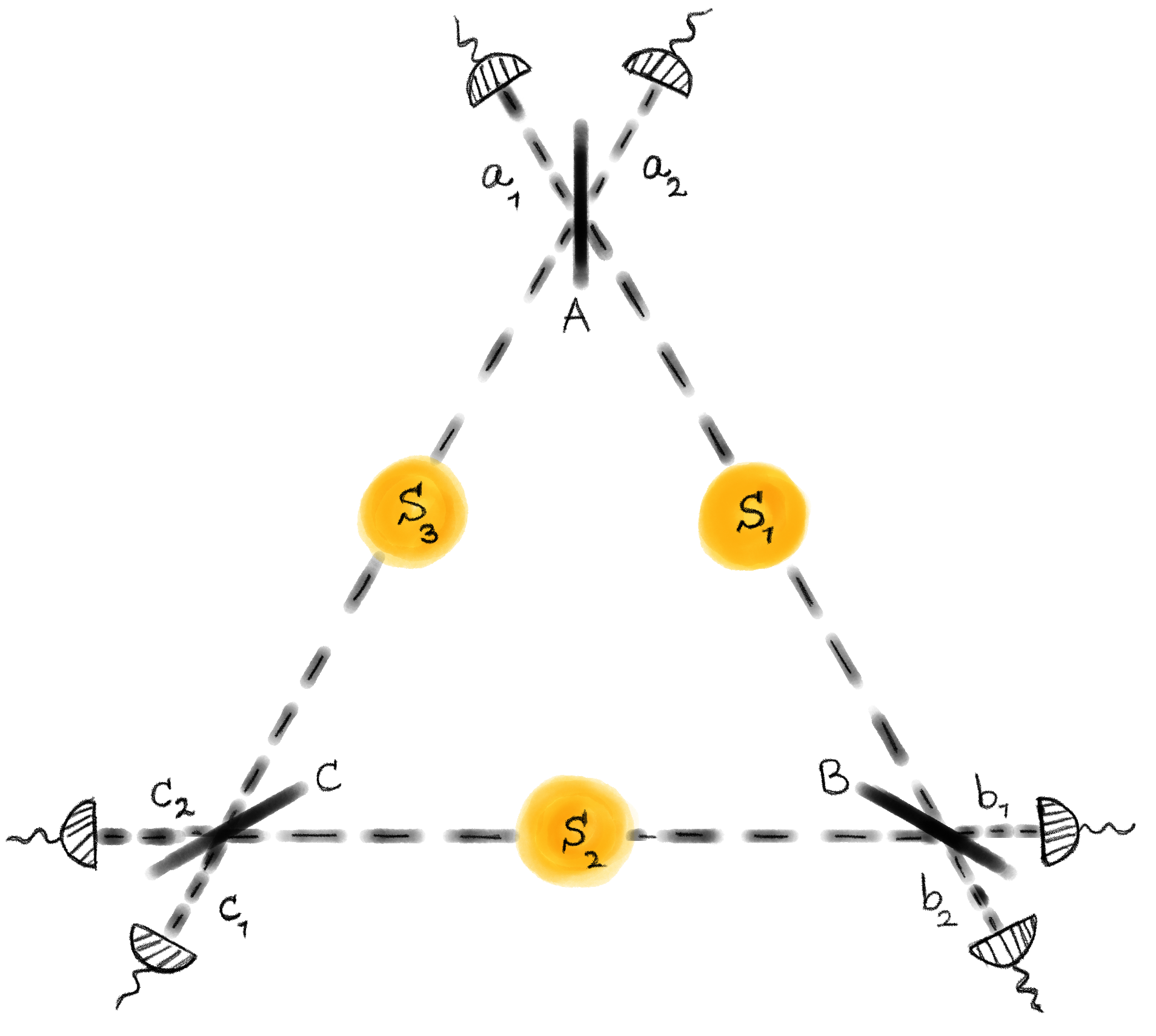}
    \caption{Three-party quantum network. In the triangular scenario with quantum sources, each source emits a single particle in a superposition of two directions of propagation towards the measuring stations at the vertices of the triangle. At each measuring station, two incoming beams are superimposed on a $50/50$ beam splitter, behind which two detectors are placed, giving two possible results. We postselect the cases where only one particle is detected at each station. }
    \label{fig:threeparties}
\end{figure}

In Fig. \ref{fig:threeparties}, we depict a three-party quantum network \footnote{For the sake of completeness, we demonstrate in Appendix A that no two-party scenario can distinguish between correlations originating from classical sources and those exhibiting quantum coherence.} where each pair of parties shares an independent source of single quantum particles that can however be prepared in coherent superposition of both paths leading to the adjacent parties. The initial state is thus represented as:
\begin{align}
\ket{\psi_{\text{in}}}=\left(\frac{1}{\sqrt{2}}\right)^3(S_{1A}^\dagger+S_{1B}^\dagger)(S_{2B}^\dagger+S_{2C}^\dagger)(S_{3C}^\dagger+S_{3A}^\dagger)\ket{0},
\end{align}
where $S_{1A}^\dagger$ (resp. $S_{1B}^\dagger$ ) is the bosonic creator operator of the mode leading to Alice (resp. Bob) of the source $S_1$, acting on the vacuum state $\ket{0}$. The same notation holds for the other sources $S_2$ and $S_3$. 

Upon crossing the paths through $50/50$ beam-splitters as in Fig. \ref{fig:threeparties}, we obtain:
\begin{align}
\ket{\psi_{\text{BS}}}=\left(\frac{1}{2}\right)^3&(a_1^\dagger+a_2^\dagger+b_1^\dagger-b_2^\dagger)(b_1^\dagger+b_2^\dagger+c_1^\dagger-c_2^\dagger)\nonumber\\&(c_1^\dagger+c_2^\dagger+a_1^\dagger-a_2^\dagger)\ket{0}.
\end{align}
Where $a^\dagger_1$ and $a^\dagger_2$ are the creation operators of bosonic modes created after the beam splitter at Alice's station, on the left ($1$) and right ($2$) sides of the beam splitter, respectively. Similarly, $b^\dagger$ and $c^\dagger$ denote corresponding creation operators at Bob's and Clair's stations. After postselecting one particle per station, we obtain the following state:
\begin{align}
\ket{\psi_{\text{PS}}}=\frac{1}{2}\left(a_1^\dagger(b_1^\dagger c_1^\dagger+b_2^\dagger c_2^\dagger)+a_2^\dagger(b_2^\dagger c_1^\dagger+b_1^\dagger c_2^\dagger)\right)\ket{0}.
\end{align}
This leads to the probability distribution:
\begin{align}
&p(a=0,b=0,c=0)=p(a=0,b=1,c=1)=\frac{1}{4},\nonumber\\&p(a=1,b=0,c=1)=p(a=1,b=1,c=0)=\frac{1}{4}.
\label{q3prob}
\end{align}
This matches the condition given in Eq. \eqref{predicate}. 
\section{An inequality to test coherence without settings}
The protocol presented in the previous sections shows that the use of quantum coherence can produce probability distributions not achievable with classical networks, even without involvement of settings, unlike normal interference experiments. An interesting question arises, if classical and quantum sets of probability distributions can be separated by inequalities, similarly to Bell's inequalities. In the appendix B we show that there is no such linear inequality to distinguish classical set from quantum one due to the following reasons: $1)$ the classical set is non convex, and $2)$ it contains all vertices of the logical polytope. 
Therefore, here we use a nonlinear inequality to distinguish classical correlations from genuine quantum correlations (in network nonlocality this is a well-known problem \cite{tavakoli2022bell}). We study the three-party case introduced above in detail and discuss the generalization to \( n \)-parties in the next section.

For three parties, the non-linear expression is as follows:
\begin{align}
    C_3=&|p(0,0,0)p(0,1,1)-p(0,1,0)p(0,0,1)+\nonumber\\&p(0,0,0)p(1,0,1)-p(1,0,0)p(0,0,1)+\nonumber\\&p(0,0,0)p(1,1,0)-p(1,0,0)p(0,1,0)+\nonumber\\&p(0,1,1)p(1,0,1)-p(1,1,1)p(0,0,1)+\nonumber\\&p(0,1,1)p(1,1,0)-p(1,1,1)p(0,1,0)+\nonumber\\&p(1,0,1)p(1,1,0)-p(1,0,0)p(1,1,1)|. \label{c3}
\end{align}
By substituting the postselected classical three-party probability distribution as:
\begin{align}
    p_{\text{cl}}(a,b,c)=&\gamma p_A(a)p_B(b)p_C(c)+ \\ &(1-\gamma)q_A(a)q_B(b)q_C(c),
\end{align}
into \(C_3\) and maximizing numerically over all possible values for \(\gamma\), \(p_A(a)\), \(p_B(b)\), \(p_C(c)\), \(q_A(a)\), \(q_B(b)\), and \(q_C(c)\), one obtains \(C_3^{\text{cl}} \leq \frac{1}{4}\). The bound \(C_3^{\text{cl}} = \frac{1}{4}\) can be achieved when the probability of any pair of the following outcomes $\{(0,0,0), (0,1,1), (1,0,1), (1,0,1)\}$is $1/2$. For instance, $p(0,0,0) = p(1,0,1) = 1/2$, which occurs when $\gamma=1/2$ and $p_A(0)=p_B(0)=p_C(0)=1$ and $q_A(1)=q_B(0)=q_C(1)=1$. However, the quantum probability distribution, as expressed in Eq.~\eqref{q3prob}, yields \(C_3^Q = 0.375\).

\section{Generalization to n-party}
\begin{figure}[ht]
    \centering
    \includegraphics[width=0.7\linewidth]{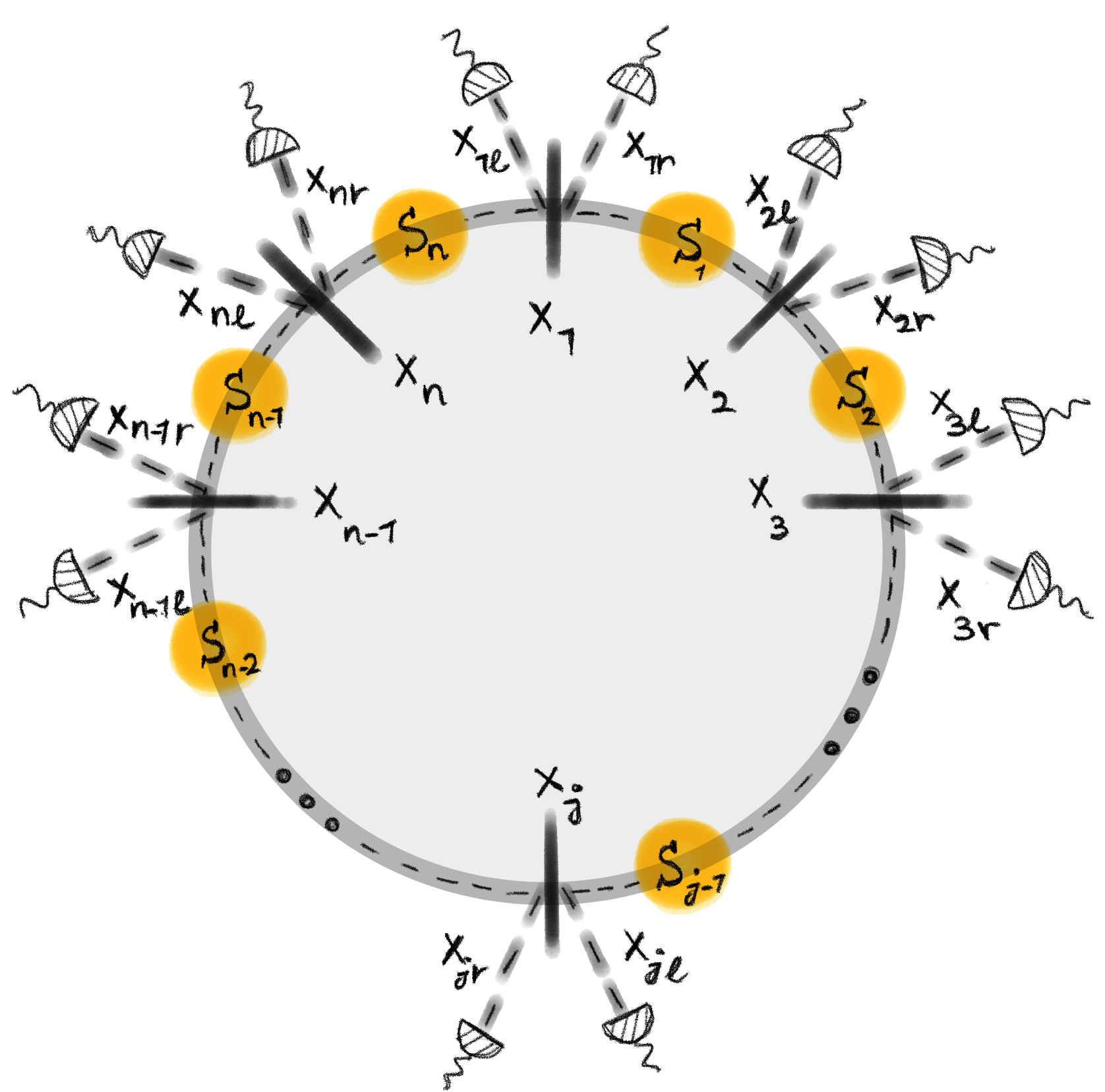}
    \caption{$n$-party quantum network. In the $n$-party scenario with quantum sources, each source emits a single particle in a superposition of two directions of propagation towards the nearest neighboring measuring stations. At measuring stations, two incoming beams are superimposed on a $50/50$ beam splitter, behind which two detectors are placed, giving two possible results. }
    \label{fig:nparties}
\end{figure}

In the general case of $n$-party network the task for both classical and quantum networks is to return the following probability distribution:
\begin{align}
    &p(a_1,a_2,\ldots,a_n|a_1\oplus a_2 \oplus \ldots \oplus a_n  = 0) \neq 0 \nonumber \ \ \& \\ &p(a_1,a_2,\ldots,a_n|a_1\oplus a_2 \oplus \ldots \oplus a_n = 1) = 0 \ \ \forall \ \ a_1,\ldots,a_n.  \label{npartiescondition}
\end{align}
Before studying the classical case, we shall first calculate the quantum correlations. 
Consider $n$-parties positioned on a $n$-partite ring network such that neighboring parties share a single-particle source, as depicted in Fig.~\ref{fig:nparties}. The input state can be expressed as follows:
\begin{align}
    \ket{\psi_{\text{in}}} =&(\frac{1}{\sqrt{2}})^n(S_{1X_1}^\dagger + S_{1X_2}^\dagger)(S_{2X_2}^\dagger + S_{2X_3}^\dagger) \dots \nonumber\\ &(S_{nX_n}^\dagger + S_{nX_1}^\dagger) \ket{0}.\label{psiinn}
\end{align}
After passing through $50/50$ beam splitters, the state becomes:
\begin{align}
    \ket{\psi_{\text{BS}}} = &(\frac{1}{2})^n(X_{1l}^\dagger - X_{1r}^\dagger + X_{2l}^\dagger + X_{2r}^\dagger) \dots \nonumber \\
    & \dots (X_{2l}^\dagger - X_{2r}^\dagger + X_{3l}^\dagger + X_{3r}^\dagger) \nonumber \\
    & \dots (X_{nl}^\dagger - X_{nr}^\dagger + X_{1l}^\dagger + X_{1r}^\dagger) \ket{0}.
\end{align}
where \( X^\dagger_{il} \) and \( X^\dagger_{ir} \) are the bosonic creation operators of modes at the \( i \)-th station, after the beam splitter, on the left and right sides, respectively. Applying post-selection on the one particle per party subspace yields:
\begin{align}
    \ket{\psi_{\text{PS}}} = & \mathcal{N}(X_{1l}^\dagger - X_{1r}^\dagger)(X_{2l}^\dagger - X_{2r}^\dagger) \dots \nonumber \\
    & \dots (X_{nl}^\dagger - X_{nr}^\dagger) \ket{0} \nonumber \\
    & + (X_{1l}^\dagger + X_{1r}^\dagger)(X_{2l}^\dagger + X_{2r}^\dagger) \dots \nonumber \\
    & \dots (X_{nl}^\dagger + X_{nr}^\dagger) \ket{0}.
\end{align}
 Where $\mathcal{N}$ is the normalisation factor. It is evident that the only non-zero terms are those with an even number of $X_{jr}^\dagger$, where $j=1,2,\dots,n$, all appearing with the same probability $p=1/2^{(n-1)}$.

On the other hand, the classical probability can be expressed as:
\begin{align}
\label{pncl}
    p_{\text{cl}}(a_1,a_2,a_3,\dots,a_n) = & \gamma \,p_1(a_1) p_2(a_2) \dots p_n(a_n) \nonumber \\
    & + (1-\gamma) \,q_1(a_1) q_2(a_2) \dots q_n(a_n),
\end{align}
where $\gamma$ is the probability of particles being distributed clockwise, and $p_j(a_i)$ ($q_j(a_i)$) is the probability that $j$th station outcomes $a_i$ given that the particles are sent in clockwise (counterclockwise) direction. This represents a convex mixture of two product distributions and thus cannot satisfy the generalized condition for given in Eq. \eqref{npartiescondition}.

\subsection{Generalization of the inequality to $n$ parties}
Given the figure of merit in Eq. \eqref{c3}, a straightforward generalization of \eqref{c3} becomes as follows:
\begin{align}
    C_n = \frac{1}{2} \Bigg| & \sum_{\substack{\vec{A} \in \{0,1\}^n \\\vec{A}^{\prime} \in \{0,1\}^n \\ \vec{A}^{\prime} \neq \vec{A} \\ \bigoplus_{i=1}^{n} a_i^{(\prime)} = 0}} 
    p(\vec{A}) p(\vec{A}^{\prime}) 
    - \sum_{\substack{\vec{B}\in \{0,1\}^n \\\vec{B}^{\prime} \in \{0,1\}^n \\ \vec{B}^{\prime} \neq \vec{B} \\ \bigoplus_{i=1}^{n} b_i^{(\prime)} = 1}} 
    p(\vec{B}) p(\vec{B}^{\prime}) \Bigg|
    \label{cn}
\end{align}

 The first summation is taken over all different vector pairs \(\vec{A}= (a_1, a_2, \ldots, a_n)\) and \(\vec{A}^{\prime} = (a_1^{\prime}, a_2^{\prime}, \ldots, a_n^{\prime})\) of length \(n\) with the parity of the components of both \(\vec{A}\) and  \(\vec{A}^{\prime}\) being zero. The second summation is taken over all different vector pairs \(\vec{B} = (b_1, b_2, \ldots, b_n)\) and \(\vec{B}^{\prime} = (b_1^{\prime}, b_2^{\prime}, \ldots, b_n^{\prime})\) of length \(n\) with the parity of the components of both \(\vec{B}\) and \(\vec{B}^{\prime}\) being one.
One can observe that the logical maximum of \( C_n \), achieved with quantum resources (Eq.~\eqref{psiinn}), is given by

\begin{align}
    C_n^Q = \binom{2^{n-1}}{2} \left( \frac{1}{2^{n-1}} \right)^2 = \frac{1}{2} - 2^{-n}.
\end{align}
We obtained classical bound \( C_n^{\text{cl}} \) by substituting \( p_{\text{cl}} \) (Eq.~\eqref{pncl}) into the expression for \( C_n \) and 
performing extensive numerical optimization. For all tested values of \( n \in \{2, \ldots, 10\} \), we consistently obtained \(\text{max}(C_n^{\text{cl}}) = \frac{1}{4} \). Consequently, we conjecture that \(\text{max}(C_n^{\text{cl}}) = \frac{1}{4}\) for all \( n \). This result implies that the discrepancy between quantum and classical correlations increases with the number of parties, reaching a maximum value of \( \frac{1}{4} \) in the limit as \( n \) approaches infinity.


\section{Conclusions}
The benefits of our investigation are twofold. Firstly, from a foundational perspective, our results provide novel insights into the understanding of quantum interference phenomena from a new angle that extends beyond standard tests. In this context, it will be interesting to see how our findings align with recent research \cite{catani2023quantum, catani2023pra} on simulating quantum interference using classical stochastic fields. Secondly, our results have implications for studying quantum interference as a resource in information processing within quantum networks. There are numerous possibilities for developing novel protocols based on quantum coherence, such as scenarios with multiple senders and a single receiver, like in coherence-based multiple access channels \cite{zhang2022building}, or development of cryptographic primitives based on single quantum particles \cite{Massa2022key}.

\section{Acknowledgements}
We thank \v Caslav Brukner, Nicolas Gisin, Sebastian Horvat, and Sadra Boreiri for useful discussions.
This research was funded in whole, or in part, by the Austrian Science Fund [10.55776/P36994] and [10.55776/COE1] and the European Union – NextGenerationEU. For open access purposes, the author(s) has applied a CC BY public copyright license to any author accepted manuscript version arising from this submission. This research was supported by the FWF (Austrian Science Fund) through an Erwin Schrödinger Fellowship (Project J 4699)


\bibliography{main.bib}

\clearpage

\section{Appendix A -- Bipartite case with no quantum advantage}
\subsection{Quantum case}

\begin{figure}[ht]
    \centering
    \includegraphics[width=0.7\linewidth]{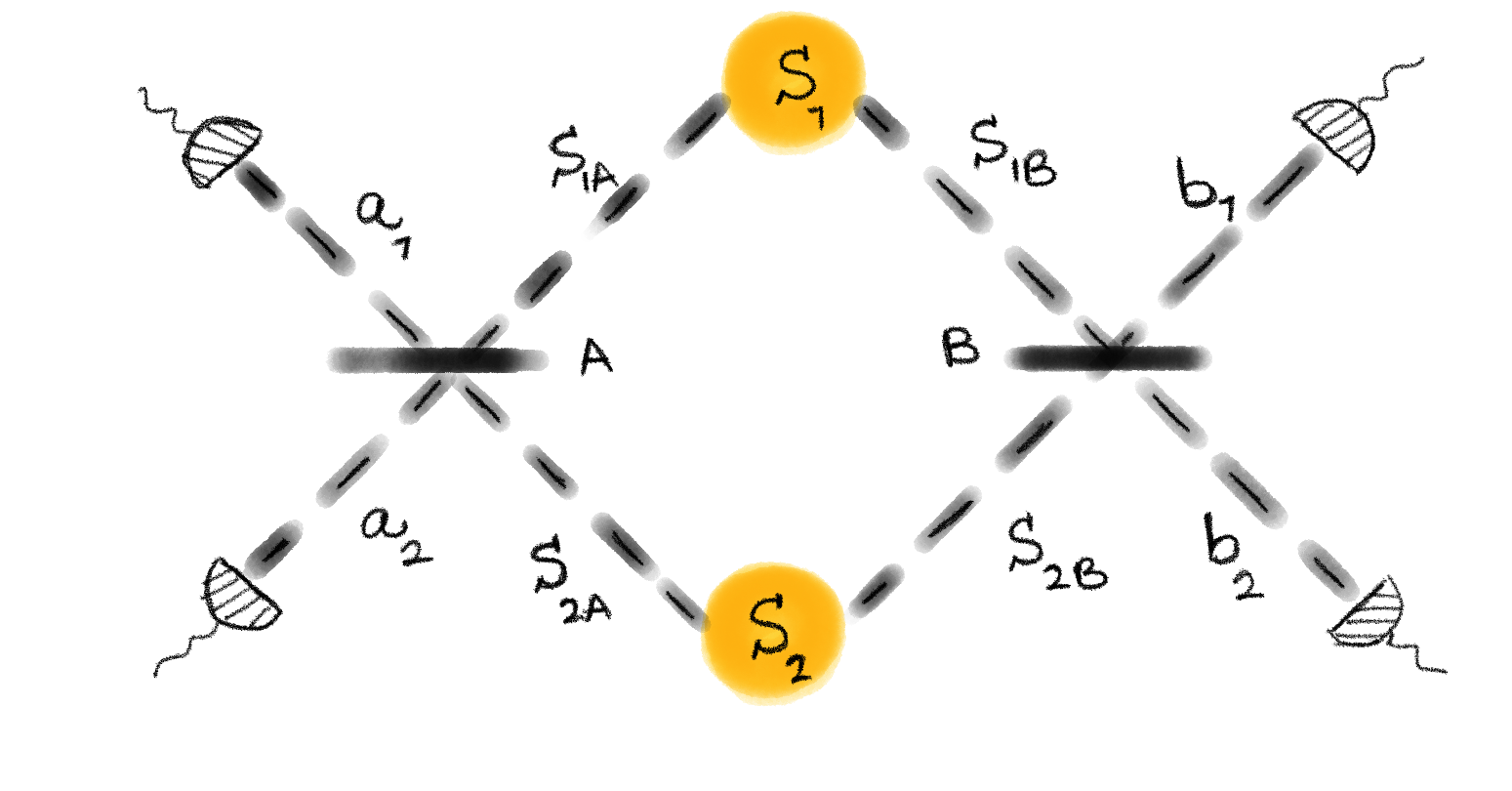}
    \caption{Two-party quantum network. In the two-party scenario with quantum sources, each source emits a single particle in a superposition of two propagation directions towards Alice's and Bob's stations. At each measuring station, the two incoming beams are superimposed on a $50/50$ beam splitter, with two detectors positioned behind it, resulting in two possible outcomes. We postselect the cases where only one particle is detected at each station.}
    \label{fig:twoparties}
\end{figure}

In the scenario illustrated in Fig. \ref{fig:twoparties}, Alice and Bob share two independent one-particle sources, labeled as $S_1$ and $S_2$. To prevent interactions between particles from different sources at the beam splitters, we selectively focus on the subspace where each party receives only one particle. Following the beam splitters, Alice and Bob conduct measurements to determine the position of the particle.

The input state can be expressed as:
\begin{align}
\ket{\psi_{in}}=(\frac{1}{\sqrt{2}})^2(S_{1A}^\dagger+S_{1B}^\dagger)(S_{2A}^\dagger+S_{2B}^\dagger)\ket{0}.
\end{align}
After the $50/50$ beam splitter, the state becomes:
\begin{align}
\ket{\psi_{BS}}=\frac{1}{4}(a_1^\dagger+a_2^\dagger+b_1^\dagger+b_2^\dagger)(a_1^\dagger-a_2^\dagger+b_1^\dagger-b_2^\dagger)\ket{0}.
\end{align}
Applying post-selection on the subspace where each party has one particle leads to:
\begin{align}
\ket{\psi_{PS}}=\frac{1}{\sqrt{2}}(a_1^\dagger b_1^\dagger-a_2^\dagger b_2^\dagger)\ket{0}.\label{sciab}
\end{align}
This state features a perfect correlation between the outcomes of Alice and Bob.
\subsection{Classical case}
\begin{figure}[ht]
    \centering
    \includegraphics[width=0.7\linewidth]{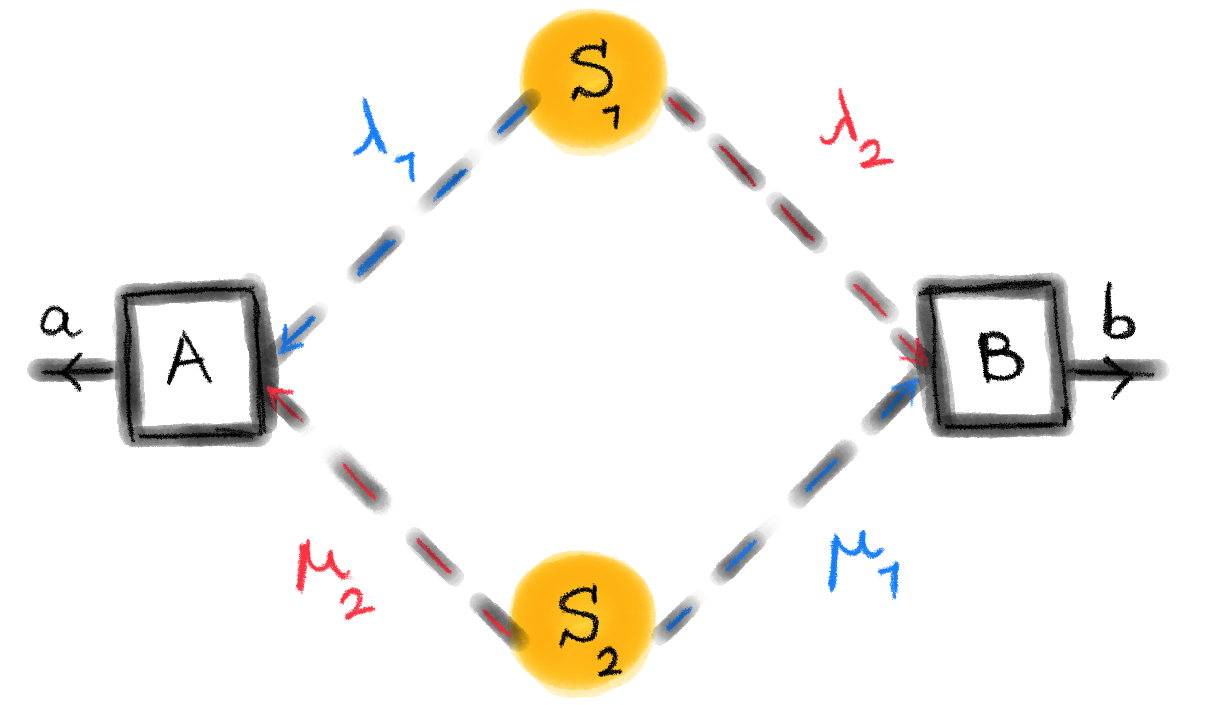}
    \caption{Two-party classical network. In the two-party scenario with classical sources, the individual particles are directed with classical probabilities to the measurement stations at the Alice's and Bob's labs. For example, source $S_1$ sends the particle to Alice's lab with probability $\lambda_1$ and to Bob's lab with probability $\lambda_2$. Since we are interested in cases where only one particle ends up in each lab, only cases where all sources send their particles clockwise or counterclockwise contribute to the correlations. }
    \label{fig:twopartiesclassical}
\end{figure}

\begin{figure}[!ht]
  \centering
       \subfloat[]{\includegraphics[width=0.4\linewidth]{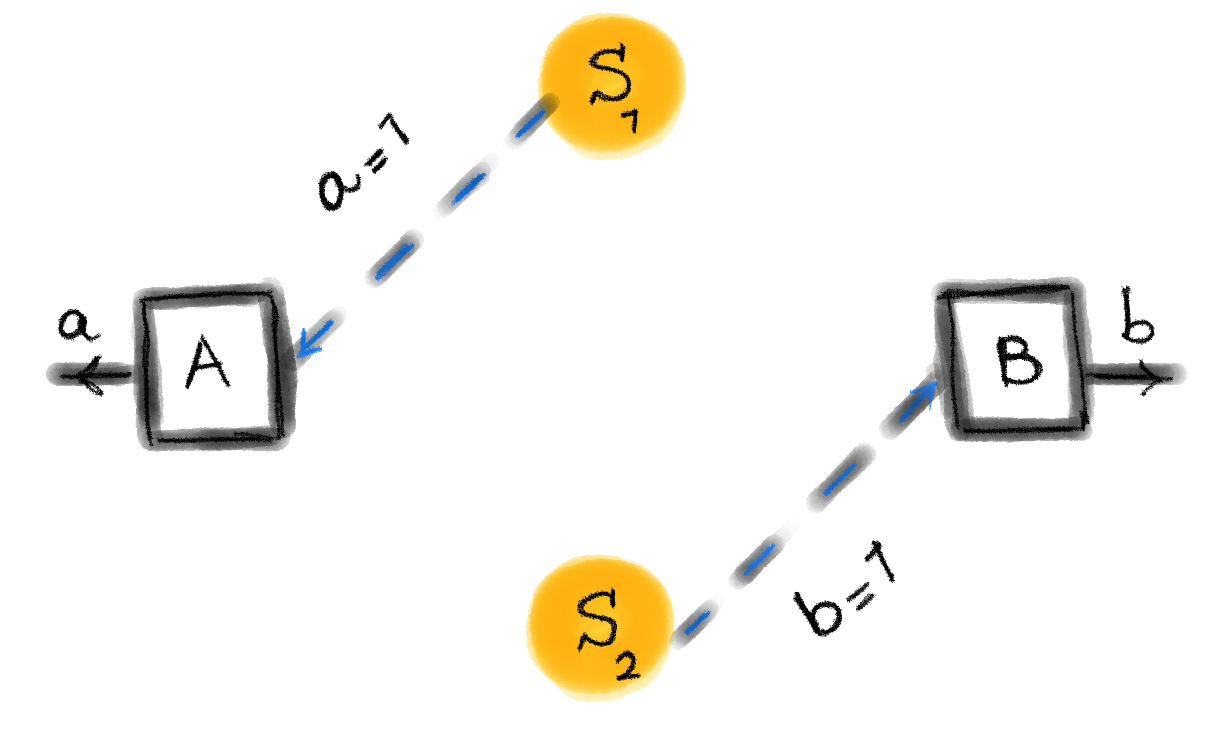}} 
         \subfloat[]{\includegraphics[width=0.4\linewidth]{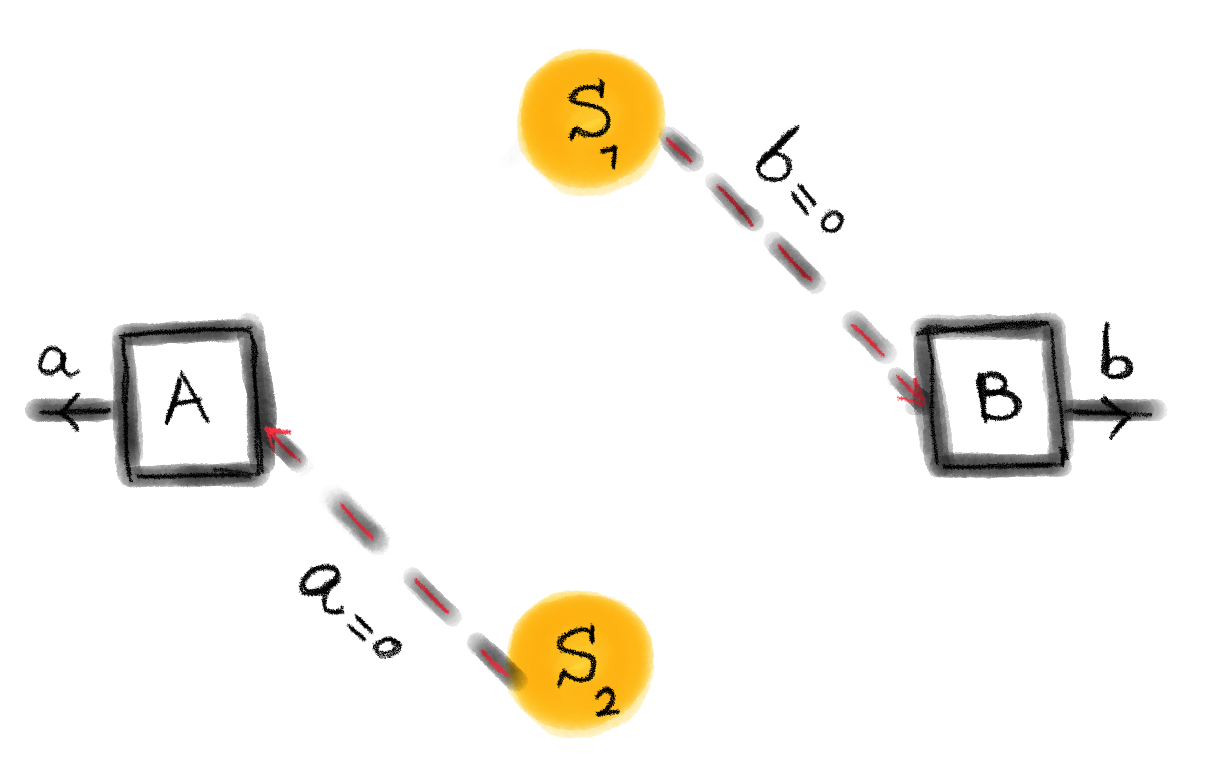}}
   \caption{Diagrammatic solutions. Perfect correlation between the outcomes of Alice's and Bob's measurements is achieved with the following classical sources: In panel (a), Source $S_1$ sends a particle with $a=1$ to Alice's lab, and Source $S_2$ sends a particle with $b=1$ to Bob's lab. In panel (b), Source $S_2$ sends a particle with $a=0$ to Alice's lab, and Source $S_1$ sends a particle with $b=0$ to Bob's lab. Each of these cases occurs with a probability of $\frac{1}{2}$.}
    \label{fig:Diagramatic solutions}
\end{figure}

In the classical scenario, illustrated in Fig. \ref{fig:twopartiesclassical}, very similar to later quantum case, Alice and Bob share two independent one-particle sources, labeled $S_1$ and $S_2$. Each particle is sent from $S_1$ ($S_2$) to Alice with probability $\lambda_1$ ($\mu_2$) or to Bob with probability $\lambda_2=1-\lambda_1$ ($\mu_1=1-\mu_2$). To prevent any interaction between particles from different sources that could alter their hidden variables, we select the subspace where each party receives only one particle. Throughout the experiment, Alice and Bob perform fixed measurements, yielding outcomes $a$ and $b$, respectively. 
The joint probability of Alice and Bob's outcomes is given by:
\begin{align}
p(a,b)=&\left(\frac{\lambda_1\mu_1}{\pi_{11}}p_A(a|S_1)p_B(b|S_2)+\frac{\lambda_2\mu_2}{\pi_{11}}p_A(a|S_2)p_B(b|S_1)\right).
\label{clprob2}
\end{align}
Here, $\pi_{11}=\lambda_1\mu_1+\lambda_2\mu_2$ represents the probability of having one particle per party, and $p_{A}(a|S_1)$ is the probability that Alice obtains result $a$, given that the particle came from $S_1$. Assuming perfect correlations such that $p(a=0,b=0)=p(a=1,b=1)=1/2$, and $p(a=0,b=1)=p(a=1,b=0)=0$, the following conditions arise:
\begin{align}
&(i)&p_A(a=0|S_1)=0 \,\,\,\&\,\,\, p_B(b=1|S_1)=0,\nonumber\\
&(ii)&p_A(a=0|S_2)=0 \,\,\,\&\,\,\, p_B(b=1|S_2)=0,\nonumber\\
&(iii)&p_A(a=1|S_1)=0 \,\,\,\&\,\,\, p_B(b=0|S_1)=0,\nonumber\\
&(iv)&p_A(a=1|S_2)=0 \,\,\,\&\,\,\, p_B(b=0|S_2)=0.\nonumber\\
\end{align}
Among these, only the combinations of $(i)\,\&\,(iii)$ and $(iii)\,\& \,(iv)$ with $\lambda_1=\mu_1=1/2$ result in correct marginals takes from state \ref{sciab}. For a diagrammatic solution, refer to Fig. \ref{fig:Diagramatic solutions}. 

In analogy to the three-party inequality introduced in Eq.\eqref{c3} in the main text, here we introduce the nonlinear inequality for two parties as following 

\begin{align}
    C_2=|p(0,0)p(1,1)-p(1,0)p(0,1)|.
\end{align}
The joint classical probability distribution (see Eq.~\eqref{clprob2}) can be represented as 
\begin{align}
    p_{cl}(a,b)=\gamma p_A(a)p_B(b)+(1-\gamma)q_A(a)q_B(b).
\end{align}
Here, $\gamma$ represents the probability of particles being distributed among the parties in a clockwise direction and $p_A(a)$ ($q_A(a)$) is the probability that Alice measures outcome $a$ given that particles has been distributed in the clockwise (counter clockwise) direction. The maximum value of $C_2=1/4$ can be achieved through perfect correlation, a condition achievable in two-party classical networks. We conclude that perfect correlations between the outcomes of Alice and Bob can be achieved by both classical and quantum networks. This is expected because all the hidden variables from both sources are available at each measurement station, thus negating any independence condition in the situation of two-party network.\\
\section{Appendix B -- Properties of the Classical Set}

The classical subset of joint probability distributions with three variables is defined by:
\begin{align}
    p_{\text{cl}}(a,b,c) =& \gamma p_A(a)p_B(b)p_C(c) + \\ 
    & (1-\gamma)q_A(a)q_B(b)q_C(c),\nonumber
\end{align}
where $\gamma$ is a probability (i.e., $0 \leq \gamma \leq 1$), and $p_A(a), p_B(b), p_C(c), q_A(a), q_B(b), q_C(c)$ are all valid probability distributions over the binary variables $a, b, c \in \{0,1\}$. 

To check whether the set is convex or not, we assume $p^1_{\text{cl}}(a,b,c)$ and $p^2_{\text{cl}}(a,b,c)$ are elements of the classical subset. Next we need to check whether $\lambda p^1_{\text{cl}}(a,b,c) + (1-\lambda)p^2_{\text{cl}}(a,b,c)$, where $\lambda$ is a probability, can be written as a convex mixture of two three-variable product distributions. It is easy to show that this is impossible, which proves that the classical set is non-convex. 
The second property of the classical subset we are interested in is whether it contains all vertices of the logical polytope of three-variable probability distributions. The vertices correspond to extreme cases where $p(a=i,b=j,c=k)=1$ for specific values of $i,j,k$, and $p(a,b,c)=0$ for $(a,b,c)\neq (i,j,k)$. These extreme cases can be attained by setting $\gamma=1$, $p_A(a=i)=1$, $p_B(b=j)=1$, and $p_C(c=k)=1$. Therefore, the classical subset contains all vertices of the logical polytope of three-variable probability distributions. As a result, the only hyperplanes that intersect the classical subset are the facets of the logical polytope. Therefore, there is no hyperplane that can distinguish classical correlations from quantum correlations.
\end{document}